\documentclass[journal=jctcce,manuscript=article]{achemso}

\usepackage[version=3]{mhchem} 
\usepackage[T1]{fontenc}       

\usepackage{graphicx,xcolor}
\usepackage{amssymb,amsmath}
\usepackage{dcolumn}    
\usepackage{rotating}
\usepackage{import}
\usepackage{color}
\usepackage{transparent}

\usepackage{dcolumn}    
\usepackage{bm}         
\usepackage{rotating}
\usepackage{color}

\usepackage{subcaption}
\usepackage{adjustbox} 
\usepackage{siunitx}
\usepackage{longtable}
\usepackage{lscape}

\usepackage{hyperref}
\hypersetup{
    colorlinks,
    citecolor=blue,
    filecolor=black,
    linkcolor=black,
    urlcolor=blue
}

\author{Judith B. Rommel}
\affiliation[University of Cambridge]
{Department of Chemistry, University of Cambridge, \\Lensfield Road, Cambridge, CB21EW, UK.}
\email{jbr36@cam.ac.uk}
\phone{+44 (0)1223 763872}

\title[Future of Computational Chemistry]{From Prescriptive to Predictive: an Interdisciplinary Perspective on the Future of Computational Chemistry} 

\abbreviations{DFT, CC, QSAR, QSPR, SE, TISE, TIDE, HF, PDE, CI, DMRG, MD, PCE}

\begin{document}
\begin{tocentry}

\includegraphics{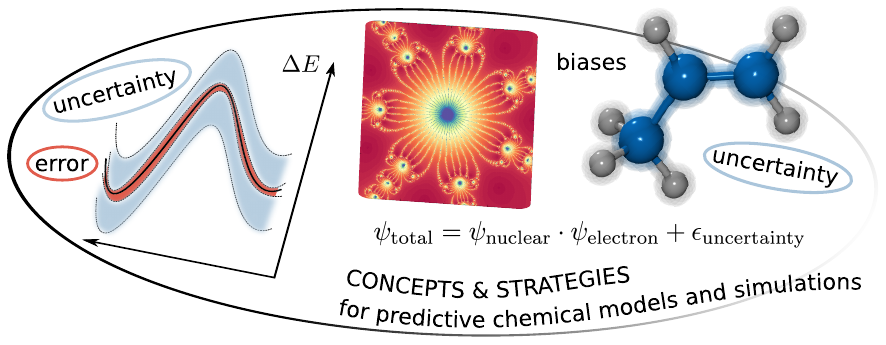}

\end{tocentry}

\begin{abstract}

Reliable predictions of the behaviour of chemical systems are essential across many industries, from nanoscale engineering over validation of advanced materials to nanotoxicity assessment in health and medicine.
For the future we therefore envision a paradigm shift for the design of chemical simulations across all length scales from a prescriptive to a predictive and quantitative science. This paper presents an integrative perspective about the state-of-the-art of modelling in computational and theoretical chemistry with examples from data- and equation-based models. Extension to include reliable risk assessments and quality control are discussed.
To specify and broaden the concept of chemical accuracy in the design cycle of reliable and robust molecular simulations the fields of computational chemistry, physics, mathematics, visualisation science, and engineering are bridged. Methods from electronic structure calculations serve as examples to explain how uncertainties arise: through assumed mechanisms in form of equations, model parameters, algorithms, and numerical implementations. We provide a full classification of uncertainties throughout the chemical modelling cycle and discuss how the associated risks can be mitigated.
Further, we apply our statements to molecular dynamics and partial differential equations based approaches. An overview of methods from numerical mathematics and statistics provides strategies to analyse risks and potential errors in the design of new materials and compounds. We also touch on methods for validation and verification.
In the conclusion we address cross-disciplinary open challenges. In future the quantitative analysis of where simulations and their prognosis fail will open doors towards predictive materials engineering and chemical modelling. 

\end{abstract}


\tableofcontents

\section{Introduction}

Computer simulations play an important role in improving and understanding of molecular science across many industries, like airplane design, screening for drugs, or materials modelling.
They provide safe test environments with reduced costs compared to traditional experimental techniques.
In both areas, in drug design as well as in airplane engineering, a major concern is to find and fix dangerous effects like design problems, which could make the airplane crash, due to the fatigue of materials on the molecular scale or toxicity, which could both harm the health of patients.
The decisions over which predictions from computer simulations are true and can be trusted are in both cases closely linked to the responsibility for the security of other people's life.
Moreover, with materials science reaching the limit of one atom layer thin materials like graphene, which exhibits surprising properties, approaches from computational chemistry - especially electronic structure theory - become more and more important in materials engineering.\cite{Heine2014,Deglmann2015} 

The accurate \emph{in silico} chemical risk assessment of drugs and nanoscale materials would enormously enhance the discovery process
and the safety design of materials and drugs. Many challenges in the development of efficient and robust tools to support (industry) scientists with reliable risk assessments in chemical or molecular systems are still open.\cite{Valerio2009,Gleeson2012,Vlachos2012,Huang2016}
Simulations of chemical systems, if based on a good model, are viable alternatives to experiments in cases where hypothesis probing is difficult, due to limitations of measurement devices or unaffordable equipments. The questions of how to know whether, why, or when a model or simulation method is good enough and how to quantify when or where they fail are tough to answer.

In this paper we will discuss where the interdisciplinary field of computational chemistry stands in terms of qualitative and quantitative concepts and approaches. Similar to airplane construction, where initially brave designers tested their guesses of what could fly themselves by risking their lives in a trial-and-error development cycle,\cite{Woltosz2012} the field can develop high level computational engineering processes involving iterations through many designs before any attempts of synthetisation are made.
We will draw connections to other fields, ranging from engineering over mathematics to visualisation science and what can be learned from their strategies, in particular about uncertainties and error estimation for risk assessments in simulations.
At the end a full picture of how to reach for an increased contribution of computational chemistry to technological development will be given.

In Sec.~\ref{sec:whatpred}, we will define the meaning of predictive simulations and extend the concepts of accuracy and uncertainty in the context of chemical simulations. 
At first the iterative cycle of general computational modelling is discussed, Sec.~\ref{sec:modcyc}, followed by a more detailed description of how data-based models fit into the cycle, Sec.~\ref{sec:databasMOD}.
In Sec.~\ref{sec:QualJudge}, the focus lies on common concepts for judging the quality of simulations of chemical systems. The full picture of uncertainties, given in Sec.~\ref{sec:uncertclass}, raises the need of extensions to these concepts in chemical modelling, model improvement, and judgement of simulation results.
Moreover, a clear distinction between creating a model, i.e., modelling, and choosing a numerical simulation method is introduced.
The model cycles of the non-relativistic Schr\"{o}dinger equation (SE)\cite{Schroedinger1926} and Density-Functional Theory (DFT)\cite{Hohenberg1964,Kohn1965} methods serve as examples in Sec.s~\ref{sec:equbasedSE} and~\ref{sec:equbasedDFT}. 
In Sec.~\ref{sec:uncertquant}, we present a broad overview of techniques, including strategies from engineering and visualisation science, to quantify and deal with uncertainties. 
Concepts and requirements for practical approaches to deal with numerical errors and chaotic behaviour are mentioned in Sec.~\ref{sec:nummath}.
The discussion then leads to standard validation and verification techniques, see Sec.~\ref{sec:valiveri}.  
The final section is dedicated to the impactful future of computational chemistry in predictive materials engineering and chemical modelling. We finish with a summary of cross-disciplinary open challenges to reach reliable risk assessments and quality control in chemical simulations.

\section{Being Predictive: from Models to Numerical Simulations}
\label{sec:whatpred}

\subsection{General Aspects of the Modelling Cycle}
\label{sec:modcyc}

The first essential step in research is to identify questions about the unknown, for which answers are sought for, often through close communication of experimental and theoretical chemists. These questions clarify what kind of data are needed and relevant for the investigation. 
The unknown often refers to target designs of drugs and materials or to behaviour at some unobserved instant in time or space.\cite{Ghanem2006} To predict new designs and unobserved behaviour means to extrapolate from observations by building at least one model.
Building models is an essential and inseparable part of scientific activity, which intends to make a specific part or characteristic of the world easier to comprehend, define, quantify, visualise, or simulate by referencing it to existing and often generally accepted knowledge or familiar mechanisms.\cite{Deglmann2015}  

A good scientific model is grounded in solid, scientific principles and systematically reproduces the physical or chemical properties of a system.
To be accepted models need testing against the complex real world, for example by comparing a \mbox{(null-)hypothesis} with experiments.
While decision-making benefits from anticipating the future, the value of the associated inference  is 
limited by the confidence in the anticipation.
Therefore, a good model, unlike number crunching, includes reliable and clear actionable outcomes with guidance of how much the prediction can be trusted when making a decision. A good model summarises evidence beyond its abstract mathematical reality, is comprehensible to other scientists,\cite{Box1976} and complements the design of experiments. 

\begin{figure}[h]
\centering
\includegraphics{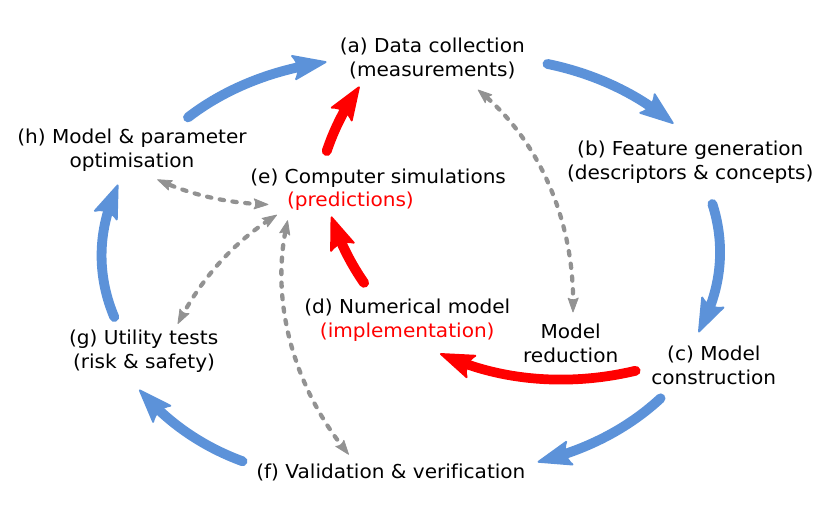}
\caption{Standard modelling cycle (blue) and the role of numerical simulations (red).}
\label{fig:modcycletot}
\end{figure}

A standard modelling cycle, see Fig.~\ref{fig:modcycletot}, consists of data collection, generating a description of features the model should include, model construction, validation of the model, practical utility tests including risk and safety assessments, model (parameter) optimisation.
Computer simulations are a central link in the modelling cycle in terms of representing the numerical implementation of a model.

A model is a physical, mathematical, symbolical, or verbal representation of a concept which has been found in order to clarify the understanding of something. 
A theory, on the other hand, is a conceptual framework of an idea, often a generalised phenomenon.
The main difference between model and theory is that theories can be considered as answers to various scientific problems while models can be considered as representations created to explain a theory.
Accordingly, chemical modelling is more than reaching agreement with experiments in the results of computer simulations. It includes thinking about concepts,\cite{Thiel2011} assumptions about the physical or chemical reality of the systems to describe, and how insights rather than just numbers can be obtained.

Chemical systems are inherently complex. Complexity can destroy the possibility of understanding a system which is the ultimate aim of chemical simulations and as such generates risk.
 Complexity can refer to the difficulty to solve a problem, the richness in terms of interconnections between subcomponents of a system, or the incomprehensibility of a huge amount of data. Obviously, the scale at which a system is observed and the practical goals of modelling it are crucial in classifying a system as simple or complex.\cite{Kevrekidis2004}
 Strategies to deal with complexity in simulations include tests with the aim to gain understanding from outside the system (risk assessment) or adding approaches which make extra information inside the system available (error and uncertainty estimates). 
 
If the mathematical description of a model is too complex to be simulated computationally then model reduction comes into play.  Model reduction for a numerical implementation is distinct from formulating a mathematical expression from a theory. Since model reduction is a process happening after the original formulation of the model it can be viewed either as part of the model construction, Fig.~\ref{fig:modcycletot} (c), or as part of the numerical implementation of the model , Fig.~\ref{fig:modcycletot} (d),
which can be iteratively improved. 
In the context of electronic structure calculations, the iterative improvement of reduced equations often refers to corrections of \emph{ab initio} or 'first principles' predictions to reach better agreements with experiments.
These practices limit the understanding and predictiveness of electronic structure simulations since they only focus on one part of the modelling cycle in their currently used forms. We will extend the related concepts in Sec.~\ref{sec:uncertclass}.

All scientific models include assumptions about reality to simplify complexity.
Thus, models are always imperfect representations of the real world and building a model is a never-ending process of discovery and refinement, rather than finding the truth.
If our assumptions are correct than stronger assumptions mean less uncertainty.
The quality of the interpretation of simulation outcomes relies on (1) the clear and precise knowledge of the assumptions used to describe the underlying physics and chemistry of a problem and (2) on the comprehensibility of the predicted results. 
Some assumptions and predictions are only valid in a certain window in time or space, under specific experimental conditions, or for certain chemical elements.
A widely used assumption is that coupled cluster (CC)\cite{Bartlett2007} is a 'higher level' theory for electronic structure simulations which results in better accuracy than Density-Functional Theory (DFT).\cite{Hohenberg1964,Kohn1965} 
Based on this assumption several studies optimised the geometries of molecules with DFT and then performed local (vibrational) analysis with CC without any re-optimisations. However, the visualisation of the underlying energy landscapes showed two distinct manifolds whose energy minima had no overlap and, thus, the predicted frequencies, despite having been performed with a 'higher level' of theory, were meaningless to understand the underlying physics of a chemical reaction.\cite{Viegas2014} 
Another example is the computational and experimental investigation of an organic multistep reaction in solution. The experimentalists concluded that the simulations were so flawed that they were 'not even wrong', because the theoretical results predicted a proton-shuttle pathway instead of the experimentally known acid-base process.\cite{Plata2015} From a modelling perspective the study clearly shows (1) that the current standard computational methods should not and cannot be used to predict certain features of solution reactions, (2) that the plethora of choice of functionals in DFT can hinder the formulation of meaningful predictions, and (3) that without the clear communication of assumptions and simplifications of a chosen chemical model the quality and reliability of simulation results cannot be ensured.

Uncertainty is intrinsic to the process of finding out what is unknown and not a weakness to avoid. 
Therefore, a prediction is complete when given with uncertainty and error bars specified by probabilities or clear quantification.
Uncertainty is present in all decisions we face. And a key risk minimisation strategy is to systematically search for uncertainties with a mixture of strategies. The benefits of uncertainty quantifications are: clear risk management, easier risk minimisation, increase in reproducibility, and improvement of the decision making process with clear distinctions between reliable and discardable results.
In many industrial applications the trade-off between obtaining predictions as quick as possible and the computational efforts of detailed simulations methods and models can be more rigorously balanced when tools and approaches are available that quantify the trade-offs in both choices. At the same time a more appropriate use of data and models is possible leading to better experimental designs and, potentially, the enhancement or replacement of high error experiments with simulations while saving the costs for expensive double explorations. Additional simulations and measurements can be planned such that the most serious uncertainties are reduced with minimal effort, for example through parametric sensitivity analysis or specific model refinement.\cite{Karniadakis2006,Walz2015}

The power of high performance computers allows us to design new models by finding new connections and interpretations directly from a deluge of measured data which go far beyond our imaginations. These models are called data-based models in contrast to models and theories which involve explicit mathematical equations and concepts. 
In Sec.~\ref{sec:equbasedSE}, we will consider the non-relativistic Schr\"{o}dinger Equation as a model based on the concept of wave functions to represent a quantum theory. 'First principle' methods are approaches in computational chemistry where explicit mathematical equations without or only a few free parameters, e.g., in Newton's equation of motion or the SE,\cite{Schroedinger1926} are available. 
We use 'first principles' in quotation marks, since many of the models used in electronic structure simulations make additional assumptions when formulating the reduced model equations, often based on empirical special case observations for a certain type of chemical system.
'First principles' model equations are representations of theories which were formulated based on former experiments, observations, and data. Many physical constants, e.g., Planck's or Boltzmann's constant, which serve as input parameters for the 'first principles' descriptions are based on measurements performed long ago. Therefore, we refer to 'first principle' methods as equation-based models.

\subsubsection{Data-based Modelling Cycles in Computational Chemistry}
\label{sec:databasMOD}

Theoretical concepts like different types of chemical bonding, dynamics of molecules, or atomic spectra lie in the core of chemistry. The most central concept of chemistry,  still loaded with a lot of controversies, is the chemical bond.\cite{Merino2015}
In chemistry, conceptual research is often connected to the extension of available visualisations, which is useful for the systematic analysis of new tools and models. For example, excited state and charge transfer simulations of larger systems in the algebraic diagrammatic construction (ADC) scheme\cite{Dreuw2015} led to the concept of state-averaged natural transition orbitals to understand correlation and relaxation phenomena.\cite{Plasser2014a,Plasser2014,Plasser2015} 

Choosing a concept, Fig.~\ref{fig:modcycletot}~(b), can mean picking a ball-and-stick model to represent atoms and bonds or a differential equation to fully describe the motions of nuclei and electrons. Descriptors used in data-based models are chosen based on physical or chemical properties, or atomistic structures.\cite{Nath2012} Electrotopological indices combine electronic information with molecular topology. As descriptor they provide information about connectivity without explicitly stating the molecular geometry.\cite{Weaver2013} 

Descriptor based stochastic models are used in materials science to identify unreported, however, chemically plausible compounds that could have interesting properties, e.g., thermodynamical,\cite{Gautier2015} mechanical,\cite{Vermeer2015} or electronic properties.\cite{Franceschetti1999}
In high-trough-put screening they contribute to finding molecules with a desired biological activity and possible drugs (docking),\cite{Lindgren2014, Weaver2013a,Weaver2013,Valerio2009,Moreira2015} or to build predictive models for \emph{in vivo} toxicity analysis.\cite{Huang2016}

In chemoinformatics machine learning algorithms are popular to study reaction mechanisms of enzymes.\cite{Mitchell2014}.  Data-based models for quantitative structure-activity relationship (QSAR) or quantitative structure-property relationship (QSPR) investigations are often regression or classification based models giving a mathematical relationship between chemical structures and biological activity, Fig.~\ref{fig:modcycletot}~(c).
The best practices for the development, validation, and exploitation of such models, which include reproducible descriptor values, confirmed model definition, or external validation sets independent of the model and the training sets, have been discussed.\cite{Tropsha2010}

The biggest challenge for data-based models used in compound screening is that the quality judgement of the models based on whether the predicted compounds have the desired effects, Fig.~\ref{fig:modcycletot}~(f), only takes place after the full screening when tests are implemented \emph{in vivo}.
Here, utility tests and the generation of predictions are closely tied together, Fig.~\ref{fig:modcycletot}~(e) and (g), and cannot be separated for the reuse of the model.
In addition, even if a model was found to be good, the assumptions and biases which went into the design of the model are hard to communicate. In order to communicate all assumptions and to allow for intermediate utility tests, new concepts are required for the construction of reusable and predictive computational models.

\subsubsection{Concepts to Judge Quality in Chemical Simulations}
\label{sec:QualJudge}

\begin{figure}[htb]
\centering
\includegraphics{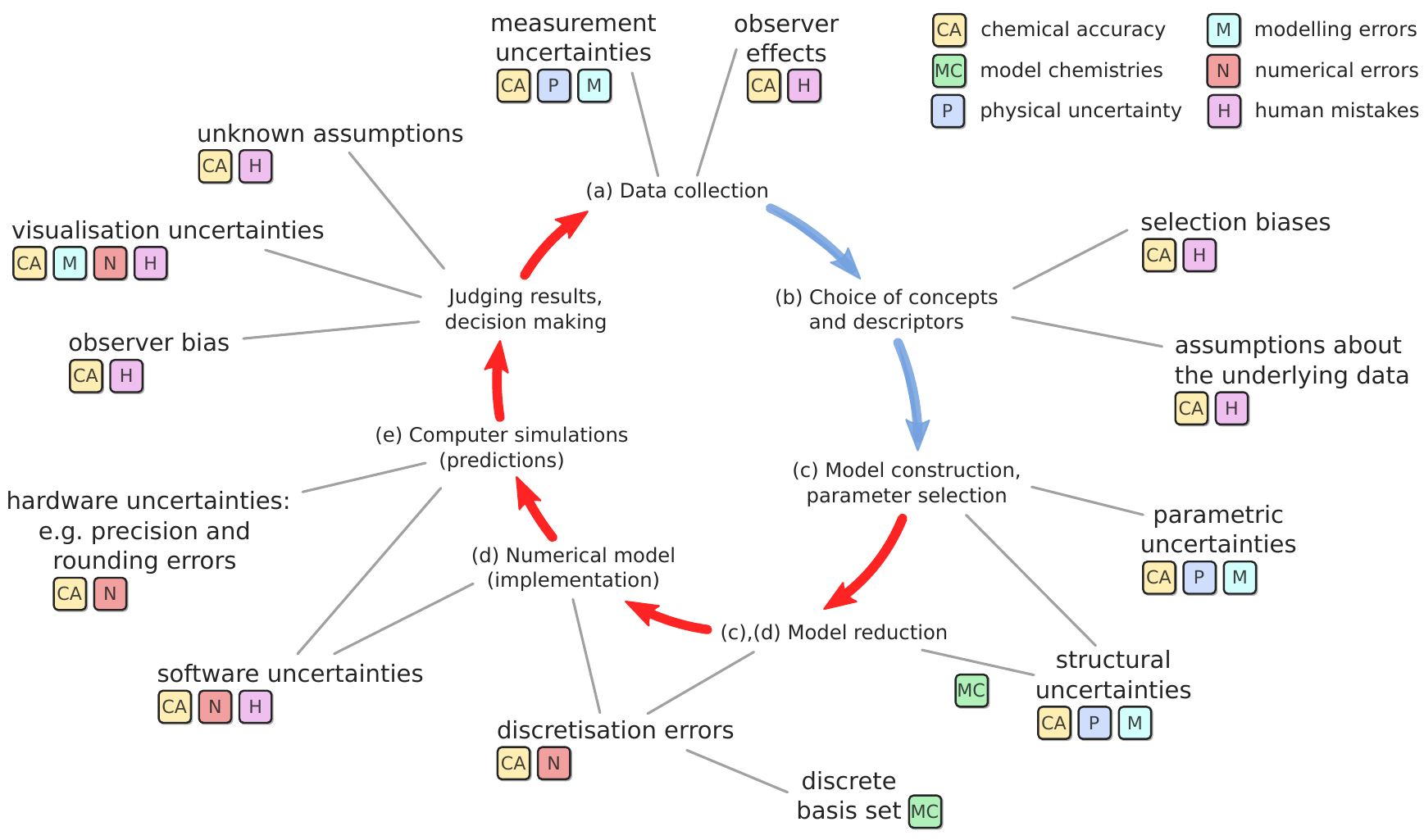}
\caption{Uncertainties in the modelling and simulation cycle.}
\label{fig:uncertainties}
\end{figure}

Computational and theoretical chemists are well aware of the fact, that in many cases one can get good results, meaning agreement with an experiment, due to hidden flaws rather than due to the properties of the underlying computational or physical models, e.g., due to error cancellation in the simulation method.
All steps in Fig.~\ref{fig:modcycletot} from measurements over formulating the mathematical model to software implementations and predictions are prone to errors and uncertainties, which in the modelling cycle are minimised in the steps of model optimisation, validation and verification, and utility tests for risk and safety assessment.
Managing the quality in all steps of the modelling cycle leads to risk reduction.
In general, challenges in judging the truth of predictions are caused by numerical errors, human mistakes, and modelling errors, see Fig.~\ref{fig:uncertainties}. 

Numerical errors include more than just precision errors. They are related to truncation of sampling spaces, round-off errors due to the finite size of memory stacks, and the stability of algorithms. These errors can be controlled for a lot of methods, see Sec.~\ref{sec:nummath}.
However, if treated without care, they can lead to the blow up of a rocket as in the example of Ariane-5 where information was lost in the data conversion from 64-bit floating point to 16-bit signed integer values due to truncations of significant numbers resulting in a series of technical failures.\cite{Lions1996}

The second source of possible unreliability of simulation outcomes are mistakes connected to humans actions, such as programming flaws, untested assumptions, person specific instead of general judging criteria, or biases in empirical or physical expert opinion.\cite{Ghanem2006}
Selection biases are present as human errors in presuppositions which are hard to control or estimate. For example the specification of a prior in Bayesian modelling\cite{Karniadakis2006} without which the simulation cannot be run. A large proportion of statistical research focuses on mathematical statistics, excluding the behaviour and processes involved in data analysis. 
Here, investigations of how people perform data analysis are required to solve the deeper problems\cite{Leek2015} and to conceptualise the process.

The third type of influence on the simulation output are modelling errors. They include the relations of the model to the physical world, its window of validity, and whether the descriptions of nature can be reproduced systematically.

In summary, the improvement of a model and its underlying approximations requires rigorous approaches in all three of the following areas: (a) the numerical simulation method, (b) the qualitative, and (c) the quantitative mathematical description of the physics and the chemistry of the systems of interest. 
Taking all of these areas into account separately gives a rich space of judging criteria and better access to bias reduction. 

To test the implemented scientific models thoroughly, the construction of error bars in the numerical simulations with clear distinctions of modelling errors and the errors introduced during the simulations by employing a certain numerical method is essential.

Common practice to deal with these issues in computational chemistry is to use the qualifier 'reasonable' which is neither a judgment of precision nor of accuracy. It empirically measures the balance of the two mixed with chemical reality, based on the researcher's intuition and without quantifiable measures.
Since good science is based on facts and data it is important to minimise the influence of intuition in the interpretation of the results of chemical simulations. Nevertheless, qualitative trends instead of quantitative agreements with experimental results are sometimes more useful when trying to predict the unknown, however, not in cases where safety and the knowledge of risks are crucial. In structural engineering, e.g., of airplanes and bridges, plenty of effort is spent to (a) ensure that errors and uncertainties in computational predictions can be estimated and (b) the overall quality of the calculations can be monitored by the user.
Thereby, safeguards for general strength and weaknesses of a computational approach can be provided throughout the entire modelling cycle.\cite{Oberkampf2002}

Another common concept in computational chemistry is 'chemical accuracy' which is usually defined as being quantitatively in agreement with experiments within an estimated root-mean-square deviation of 1\,kcal\,mol$^{-1}$ in binding energies\cite{Jensen2015} or, instead of matching numbers, as qualitatively reproducing the major physical and chemical properties of a system. 
In contrast, the precision of a calculation is a measure for numerical errors related to rounding of numbers due to their machine representation.\cite{Hoffmann2008}

Quality, accuracy, and robustness of computational methods have been identified as an essential prerequisite in complex electronic structure simulations such as photocatalysis or plasmonics.\cite{Honkala2013}
Quality in simulations can be ensured by making errors and uncertainties controllable and visible. Reproducibility indicates reliability and as such ensures a level of quality. 
Reproducibility of published results of computer simulations is linked to the access to implemented codes, workflows, environments, data, and experimental inputs.\cite{Walters2013}

Many computational results are starting points for further analysis, especially in multiscale simulations, and serve as input for further computations. Explicit error and uncertainty signatures can increase the trust in predictions based on reused data while clearly communicating the limitations and potential risks.

Computational chemistry is a largely empirical field whose predictions contain substantial uncertainty. And yet the use of standard statistical methods to quantify this uncertainty is often absent from published reports.\cite{Nicholls2014,Nicholls2014a}
In the next section we will give an overview about the main uncertainties which can arise in the modelling cycle and then discuss a range of possibilities to deal with them.

\subsection{Classifications of Uncertainties and Errors}
\label{sec:uncertclass}

The uncertainty of a prediction has at least two dimensions: (1) a context dependent quantification of the error in predictions and (2) to what extent we have trust in the model to make the intended prediction.\cite{Sahlin2015}
Risk analysts distinguish between reducible and irreducible sources of uncertainty.
Reducible uncertainty is related to observation errors and can be reduced by improving our knowledge, e.g., through data collection. However, in realistic simulations uncertainty is irreducible beyond some level or scale, for example when mixing chemicals in a reactor there will always be unmeasurable background turbulences.
Irreducible uncertainty, includes sources of randomness that are natural or an inherent property of a response variable, and is commonly referred to as variability. 
A response variable with variability, e.g., toxicity of a chemical tested in different individuals, may take different values for the same compound under repeated observations.\cite{Sahlin2015} The presence of uncertainties in simulation technologies does not devalue the approaches if uncertainties are cautiously taken into account.

When collecting data experimentally, Fig.~\ref{fig:uncertainties} (a), samples are often taken as discrete numerical quantities from continuous ranges, e.g., at discrete points in time. Since all numerical quantities can only be measured with finite precision, their values are all afflicted with some measurement uncertainty. 
Therefore, the measured results should always contain two entities: the measured value and some indication of its uncertainty, e.g., its confidence region.
Another source of uncertainty while collecting data are observer effects, due to human influence.\cite{Breznau2015}
During the generation of features (choice of concepts and descriptors) assumptions about the underlying data are made.
Selection biases contribute when experts choose chemical and physical concepts, Fig.~\ref{fig:uncertainties} (b), to describe the observed behaviour. 

In model construction, Fig.~\ref{fig:uncertainties} (c), we can distinguish structural uncertainties and parametric uncertainties.
Structural uncertainties arise from assumptions about how the physical world can be interpreted in terms of mathematical equations or in Albert Einstein's words: 'as far as propositions of mathematics refer to reality, the are not certain; and as far as they are certain, they are do not refer to reality.'\cite{Einstein1921}
Parametric uncertainties originate from the lack of variability in deterministic parameters or the limited choice in parameter selections, both providing incomplete information compared to the natural system.

The uncertainties in Fig.~\ref{fig:uncertainties} (a) to (c) are sometimes referred to as physical uncertainties which include errors due to imprecise or unknown material properties, boundary and initial conditions, equations of state, constitutive laws, etc.\cite{Karniadakis2006}
In contrast, numerical uncertainties are related to numerical implementation and simulation of a model.
Numerical uncertainties and errors arise while solving problems whose complexity requires simplifications when implementing the model numerically, Fig.~\ref{fig:uncertainties} (d). These simplifications (model reduction) can involve changes in the mathematical descriptions of the simulated system or the generation of a discrete representation of continuous equations.
Numerical uncertainties include spatial and temporal discretisation errors, errors in solvers (e.g., incomplete iterations, loss of orthogonality), geometric discretisation (e.g., linear segments), artificial boundary conditions (e.g., infinite domains, poorly known boundary conditions, missing initial conditions), etc.\cite{Karniadakis2006}
Formulating all physical concepts in purely (discrete) algebraic form so that computers can calculate them immediately is an approach to get rid of discretisation errors which is, however, due to its required effort, impractical.\cite{Tonti2014,Koren2014} 

Software uncertainties arise during software development, Fig.~\ref{fig:uncertainties} (d), for computer simulations. They are mainly impacted by inevitable human errors. 
While calculating predictions with numerical solvers and analysing these results, Fig.~\ref{fig:uncertainties} (e), uncertainties arise from numerical simulations with finite precision which lead to truncation and rounding errors.\cite{essex2000}
Also updates, changes, and choices of local cluster configurations  in high performance computing, sometimes without notifying the end users, can cause hardware uncertainties.\cite{Goddeke2015}
Similarly, there are uncertainties which can arise in the visualisation pipeline giving rise to uncertainties in the data interpretation, judgement, and decision making process.

The general modelling cycle, Fig.~\ref{fig:modcycletot}, contains steps to minimise uncertainties and risks: (f) validation and verification, discussed in Sec.~\ref{sec:valiveri}, (g) utility tests to analyse risk and safety, and (h) model or parameter optimisation.
The concept of chemical accuracy, defined in Sec.~\ref{sec:QualJudge}, comprises all uncertainties mentioned here. 
Methods, models, and theories for nanoscale engineering should explicitly account for these measurement and other uncertainties, at best separately for each of the uncertainties.
The final result of a chemical simulation depends on numerical precision, the numerical solvers, intermediate data, compilers, mathematical libraries, hardware architecture, and so forth and requires careful analysis for risk reduction and predictive results.

\subsubsection{Equation-based Modelling Cycle for Electronic Structure Simulations}
\label{sec:equbasedSE}

\begin{figure}[htb]
\centering
\includegraphics{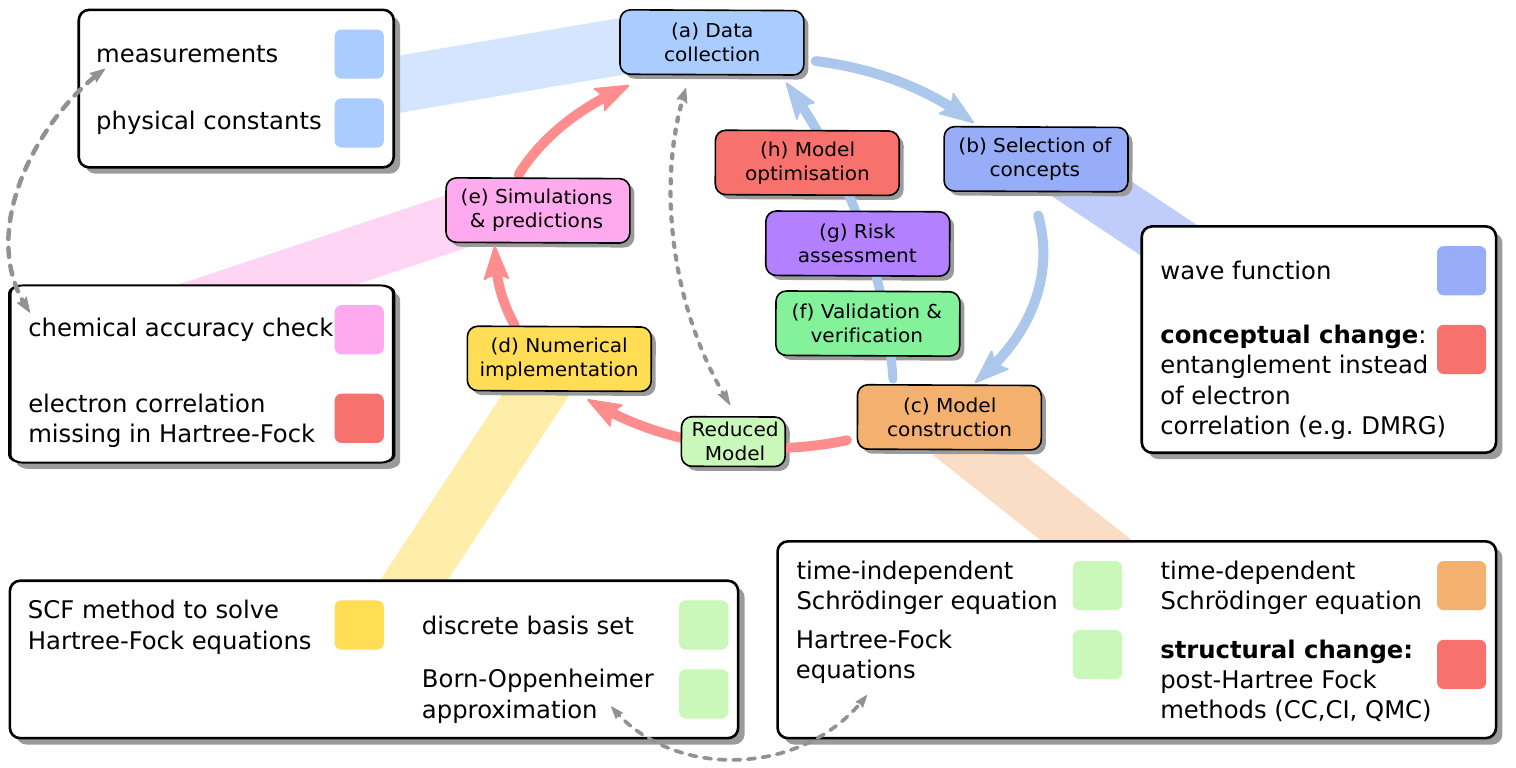}
\caption{Equation-based modelling cycle in electronic structure calculations: the time-independent SE.}
\label{fig:TISEcyc}
\end{figure}

The quantum world has many languages and concepts for its explanation, e.g., the most common Copenhagen interpretation,\cite{Kleppner2000} the modal interpretation,\cite{Bub1996} or quantum Bayesianism.\cite{Stairs2011,Timpson2008} All of them have unresolved deficiencies,\cite{ Tammaro2014} which is one of the key challenges encountered for engineering on the quantum scale.
Heisenberg's uncertainty principle, which states that either a particle's position or its momentum can be measured precisely, engraves measurement uncertainty, Fig.~\ref{fig:uncertainties} (a), into theories and models related to quantum mechanics.
Therefore, uncertainty quantification in quantum mechanical models is a difficult endeavour.

The SE is a partial differential equation (PDE) that describes how the quantum state of a system evolves over time, using electron wave functions as a central concept, Fig.~\ref{fig:TISEcyc} (b). 
Computational chemistry offers solvers for the time-dependent SE (TDSE) and the time-independent SE (TISE).
Changing the TDSE to a static TISE is a form of complexity and model reduction,  because the mathematical equations of the model are altered and simplified which impacts the physical descriptions and concepts of the chemical systems of interest, Fig.~\ref{fig:TISEcyc} (c).

Solving the non-relativistic TISE, requires the knowledge of the many-body wave function in the many-body Hilbert space, whose dimension scales exponentially with the number of particles. From here an intractable complexity arises which requires both model reduction and numerical approximations.
Therefore, many problems obtain an exact solution to an approximation of the problem rather than an approximate solution to the exact problem via model reduction.

Another form of complexity reduction to find numerical solutions for the TISE is the Born-Oppenheimer separation ansatz which assumes that the motion of electrons is instantaneous in comparison with the motion of the nuclei, since the nuclei are much heavier than the electrons. The approach is similar to the separation of variables ansatz for differential equation solvers and, despite neglecting some physical phenomena, considerable as being part of the error introduced during the numerical implementation of the model equations, Fig.~\ref{fig:TISEcyc} (d).

For elements with high speed electrons in the vicinity of heavy nuclei, which give rise to relativistic mass increase of electrons or spin-orbit interactions due to magnetic induction, TISE does not catch the essential physics involved and requires correction terms.\cite{Klopper2010,Saue2011,Cheng2012,Berger2015}
A common iterative TISE solver is the Hartree-Fock (HF)\cite{Hartree1929,Fock1930} or self-consistent field method, a mean field approach, which neglects essential many-body effects due to the simplifications of the physical model: electron correlations.
Both, relativistic effects in the TISE and electron correlations in HF, require model refinements and optimisation, Fig.~\ref{fig:TISEcyc} (h), which are achievable by introducing parametric correction terms, choosing different concepts, or alternating the model equations.
Post HF methods incorporate electron correlation; Coupled Cluster (CC)\cite{Bartlett2007} and Configuration Interaction (CI)\cite{Werner1988} approaches construct multi-electron wave functions from one-body orbitals, i.e., the model equations are extended and correction terms added. The drawback of these approaches is that they converge slowly towards the exact many-body wave functions and may require a huge amount of one-body contributions.
Quantum Monte Carlo (QMC) methods overcome these limitations by switching to a different numerical strategy and solving the multi-dimensional integrals in the many-body problem with Monte Carlo sampling. QMC flavours include variational or diffusion QMC methods\cite{Foulkes2001,Caffarel1988} and application to solid-state problems.\cite{Booth2009,Booth2013}

Another problem with CC and CI is the description of a cusp in the electron-electron interaction. Changing to a different, however similar, mathematical description and modifying the model equations by explicitly introducing the distance between two electrons in the expansions provides an approach for a better and more efficient approximation of the exact electron correlation.\cite{Ten-no2012,Kong2012}
In Density Matrix Renormalisation Group (DMRG)\cite{Szalay2015,Marti2010,Chan2011,Kurashige2013} and Matrix Product States techniques the central mathematical concepts of CC and CI are changed replacing electron correlation with entanglement.\cite{Bach2014}
The change in concepts also removes the cups problem of CC and CI. DMRG allows to simulate the electronic structure of molecules with many coupled unpaired electrons for example in the initial stages of photosynthesis.\cite{Harvey2013} It can be extended for relativistic problems\cite{Scott2015}
and is efficient for complexity reduction to reach linear scaling in computational effort with the number of nuclei.\cite{Khoromskaia2015}

In computational chemistry the concept of model chemistries states that chemical accuracy in an electronic structure calculation is achieved by a balanced combination of the chosen methods and basis sets.\cite{Hehre1986,Pople1999,Saue2011} 
Here, the word method refers to approaches for solving TISE, e.g., HF, CC, DMRG, CI, QMC, with reduced complexity due to simplifications of the physical model and the mathematical equations in preparation of the numerical implementations.
Finite basis sets, for the finite description of a wave function, are a necessity of the discrete numerical representation of quantities in computers. The truncation of basis functions is therefore a numerical error rather than a problem of the physical model, see Fig.~\ref{fig:uncertainties}.

To introduce error quantifications and uncertainty estimations we suggest a clear distinction of the types of approximations applied to both models and equations. The contributions to errors and uncertainties due to model reduction can then be optimised and balanced explicitly for increased (chemical) accuracy at similar computational costs. 
In general, we suggest to extend the concept of model chemistries and chemical accuracy with a variety of techniques and concepts to deal with as many as possible of the uncertainties and errors mentioned in the previous section to reduce risks.
The precise control of quality in each step of the modelling cycle, Figures~\ref{fig:TISEcyc} and~\ref{fig:modcycletot}, is then in close reach, moving the fields towards nanoscale engineering.

\subsubsection{Reliable and Predictive Results: Efforts in Density-Functional Theory}
\label{sec:equbasedDFT}

\begin{figure}[htb]
\centering
\includegraphics{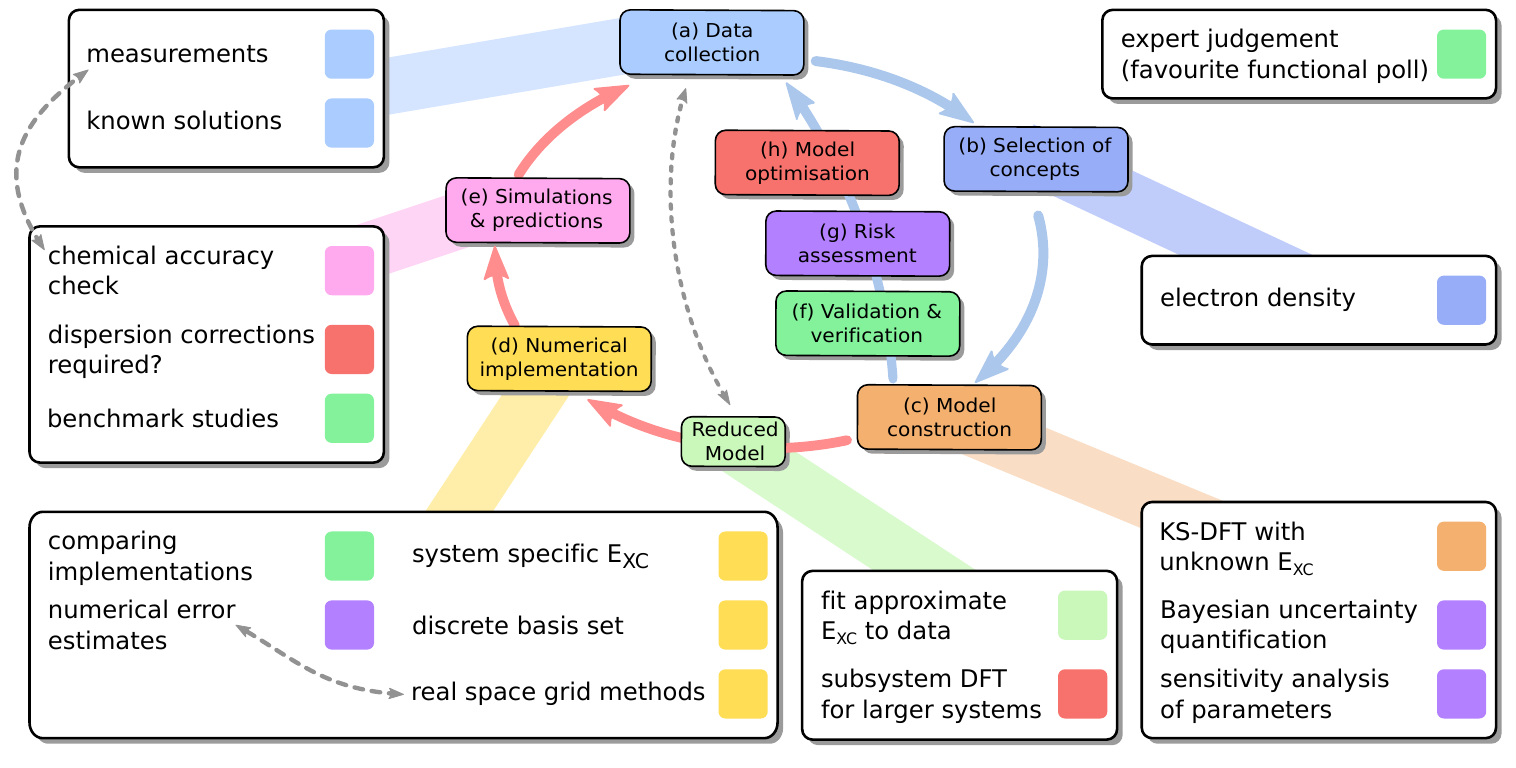}
\caption{Equation-based modelling cycle in electronic structure calculations: Kohn-Sham density-functional theory (KS-DFT).}
\label{fig:dftuncer}
\end{figure}

In Kohn-Sham DFT (KS-DFT) the concept of the ground state N-electron wave function is replaced by the mathematically simpler one-electron charge density.\cite{Hohenberg1964,Kohn1965}
In a sense the complexity of the many-body problem is reduced by mapping it onto a single-body problem including electron-electron interaction and by choosing the electron density as the central concept, Fig.~\ref{fig:dftuncer} (b).
In conceptual DFT, the electron density is the fundamental quantity for describing atomic and molecular ground states to sharply define other chemical concepts like electronegativity.\cite{Geerlings2003}
Recently, improvements in the definition of the chemical potential in conceptual DFT were suggested based on observations of the role of symmetry laws in the chemical concept of electronegativity.\cite{VonSzentpaly2015a,VonSzentpaly2015}
Here, we consider DFT (KS-DFT) as a theory and tool to calculate molecular energetics and properties, especially for systems where the TISE solvers are computationally too demanding.\cite{Jones2015}

DFT studies have successfully brought insights into experimental results: favourable reaction mechanisms in organic chemistry,\cite{ Fahrenkamp-Uppenbrink2008,Nova2014,Rommel2011b,Meisner2011,Rommel2011c,Rommel2011d,Rommel2012,Abad2014}
condensed matter physics, materials science, high-pressure and solid-state physics,\cite{Hasnip2014} and in heterogeneous catalysis.\cite{Lopez2012}

Clearly defined workflows to automate chemical modelling are becoming more and more popular in the experimental communities for example in spectroscopic analysis.\cite{Herres-Pawlis2015,Kruger2014}
Many quantum chemical workflows in condensed matter and materials science rely on DFT calculations to analyse measurements or to plan upcoming experiments.\cite{Kruger2014}
Workflows require reliable analysis and visualisation tools as well as generally accessible, quantitative, and qualitative judgement criteria for the post-processing of the simulation results.
Therefore, the question of reliability and expected error of DFT calculations becomes more and more important.

One big challenge in DFT is the choice of the case specific exchange-correlation density-functional E$_{\textrm{XC}}$ which carries the most important information of how the model describes chemical reality, inclusive model, parametric, and measurement uncertainties, Fig.~\ref{fig:dftuncer} (c) and (d).
The exact and explicit formulation of E$_{\textrm{XC}}$is unknown.\cite{Cohen2012} There is a plethora of functionals to choose from, many are fitted to data from measurements or known solutions of a wide variety of chemical problems and systems, Figures~\ref{fig:dftuncer} (a), (c), (d). The real physical reason behind a particular result is often incomprehensible, especially to newcomers in the field.\cite{Jones2015}
Choosing a functional for a specific problem independent of experiments is difficult since the  judgement process is influenced by the employed statistical measures (mean
error, mean absolute error, variance, etc.) and the biases related to them, Fig.~\ref{fig:dftuncer} (b).\cite{Jones2015} Additionally, for atomisation energies, often used to parametrise new functionals, the mean absolute error is not well defined in the limit of large molecular sizes.\cite{Savin2015}
Although many properties are well described by DFT, in some systems the modelling error (e.g. self-interaction error) can be too large to predict the reactivity of the system, if an inappropriate functional is chosen.\cite{Johansson2008} In some problem classes DFT also underestimates transition-state barriers in reactions or band gaps of materials.
The major obstacle here is to communicate all the assumptions that go into the design of each functional, the cases where a functional is good to use, the uncertainty related to the functional as well as the confidence one can have in the DFT predictions.

Models based on DFT are continuously improved, Fig.~\ref{fig:dftuncer} (h),\cite{Li2015a,Teale2016} e.g., to incorporate new concepts for excited state simulations.\cite{Corral2014,Casida2009,Casida2012}
Many people are working on improving the functional via alternative formulations\cite{Mirtschink2013, Malet2012, Kvaal2014} for better predictions of transition metals, stretched bonds, and Mott insulators.\cite{Malet2014,Perdew1981,Toher2005, Mori-Sanchez2014,Mori-Sanchez2014a}

Despite its success for biomolecular systems, DFT is unaffordable for simulations with millions of atoms due to extensive memory requirements.\cite{Hasnip2014}
Here, model reduction comes into play. An alternative to KS-DFT is subsystem DFT which reduces the complexity of the problem by partitioning the density of a system into an active and an environmental system.\cite{Jacob2014} The model equations are modified, Fig.~\ref{fig:dftuncer} (c), and the approach can reach linear scaling with the number of non-covalently bound molecules.\cite{Pavanello2013}

Methods optimising simulation grids to reduce the number of grid points in local basis functions while still knowing the accuracy of the performed calculations can improve the memory requirements. Advances in real-space numerical electronic structure methods for DFT have been reported recently.\cite{Frediani2015,ScottBobbitt2015,Nagy2015,Zuzovski2015,Natan2015,Kim2015,Lin2013,Lin2012,Kottmann2015,Tsuchida2015}

Many real-space methods are wavelet based,\cite{Mallat1989} take sparsity into account,\cite{Frediani2013} provide controls for numerical errors,\cite{Durdek2015,Bischoff2013} are robust and efficient,\cite{Genovese2015} allow for basis set truncation errors below chemical accuracy and with predetermined numerical accuracy, Fig.~\ref{fig:dftuncer} (d),\cite{Flad2015} and they are easily adapted to massively parallel simulations.\cite{Losilla2015,Jensen2014,Hu2015} Some of them scale linearly with system size\cite{Mohr2015} and they have a wide range of potential applications, including optimal control of quantum systems or simulations of plasmonic systems\cite{Andrade2015} and spectra.\cite{Natarajan2012}
Furthermore, an orbital-free stochastic approach has been suggested which allows for simultaneous, direct updates of a density matrix and, therefore, reduces computational and storage overheads.\cite{Beck2015}
Real space methods are excellent to improve the accuracy and the control of numerical errors in DFT orbital calculations. However, the problems related to the model approximations, e.g., the approximate exchange-correlation functionals, still persist.

More improvements happen at a subsequent step in the modelling workflow, the model and parameter optimisation stage. By comparing the predicted results to experimental results some chemical systems and models were found to demand dispersion corrections, i.e. correction terms after the numerical simulations have finished, Fig.~\ref{fig:dftuncer} (e). These corrections allow to reach chemical accuracy within 0.5~to~1\,kcal\,mol$^{-1}$ for chemical reactions and 0.1\,kcal\,mol$^{-1}$ for conformational energies.\cite{Schwabe2008,Grimme2011}
Dispersion corrections are usually sufficient for larger systems where hydrogen bonding is important. However, they can be quite case specific.\cite{Korth2013,Boese2015}
Further corrections are available for computed thermodynamical parameters.\cite{Hopmann2016} 
To validate and compare performance and robustness of density-functionals without uncontrolled chemical biases, special test sets have been introduced.\cite{Korth2009} Other benchmark studies looked at the chemical accuracy of functionals with respect to thermochemistry, kinetics, non-covalent interactions,\cite{Goerigk2011} or relative bond dissociation energies.\cite{Hemelsoet2011}
It was found that the chemical accuracy of the functionals increased with their theoretical complexity, in accordance with the 'Jacob's ladder' metaphor.\cite{Perdew2001}

Due to internal operations different DFT software packages which claim to use the same functionals and the same basis set can yield different absolute numbers.
Precisely these internal operations, based on various assumptions, should be incorporated in the judgement of a predicted outcome.\cite{Hoffmann2008}
A recent study showed that despite variations in the calculated values of various software packages many of them converged towards a single value, with an accuracy comparable to those of experimental measurements.\cite{Lejaeghere2016}
Software uncertainties in the numerical implementation of DFT, Fig.~\ref{fig:uncertainties}~(d) (e), were investigated by comparing several DFT codes. The differences in the results had less variation than the typical deviation of experimental measurements. Thus, the predictions were assumed to be identical.\cite{Lejaeghere2013a}

The evaluation of the reliability of DFT calculations so far mainly built on comparisons to experiments or to data sets of so called 'higher-level' calculations, meaning 'first principles' wave function based calculations.
Directly calculated properties such as bond strengths, bond lengths, or activation energies of elementary processes provided measures of the expected accuracy and uncertainty. The challenge is to relate these uncertainties to predicted complex properties like mechanical strength, phase stability, and catalytic reaction rates.\cite{Medford2014}
A clear correlation between theoretical and experimental data is key to enable reliability checks through validation and verification.
Systematic error reduction and sensitivity analysis in theoretical setups (i.e., the choice
of DFT functional, basis set, and numerical solution model) as well as in the measurement conditions provide a way to eliminate uncertainties in the comparison between experiments and calculations.\cite{Konezny2012}
Despite the exact exchange-correlation functional is unknown it is possible to obtain functionals including uncertainties with respect to experimental results. Atomisation energies of solids and small molecules served as input to a Bayesian machine learning approach.\cite{Aldegunde2016}

For precise quality control and risk analysis new concepts and approaches are required which include reliable analysis of solution stability, possible biases of the results, as well as error or uncertainty measurements throughout the entire modelling cycle. These are all desirable features to ensure scientific integrity, public accountability, and social responsibility through clear reproducibility in the conduct of science.
Chemical accuracy is simply not enough to reliably design and engineer nanoscale devices.

\section{Risk Assessment: Managing Uncertainties and Errors}
\label{sec:uncertquant}

The results of computer simulations are only of value when the amount by which they probably differ from the truth, i.e., the sum of their errors and uncertainties, is so small as to be insignificant for the purposes of the described (experimental) prediction
The most common way to represent, analyse, and deal with uncertainty is to employ methods from probability theory and statistics,
\cite{Sahlin2015,Nicholls2014,Spiegelhalter2014}
as for example in the uncertainty assessment of glass transition temperature in polymers\cite{Patrone2016} obtained from molecular dynamics (MD) simulations.\cite{Alder1959,Rahman1964}
Several strategies are equally correct.
Their choice depends on the situation and purpose of the assessments. 
At first the modeller assessing uncertainties decides which uncertainties are considered and whether the focus will lie on knowledge-based uncertainty or if variability is included.
Then the decision is about measuring and characterising uncertainty, in particular, which probability models to employ and how to interpret probability. Potential pitfalls in statistical modelling include misinterpreting correlation with causation, selection biases, oversimplifications (e.g., through regression to the mean), confirmation biases, or false syllogisms in the logical design.\cite{Spiegelhalter2014,Gelman2013,Jain2012,Leek2015}

\subsection{Uncertainty Quantification}
The following sections provide an overview of strategies to assess uncertainty including modelling (Bayesian approaches), sampling (bootstrapping), or post modelling (parametric sensitivity analysis) of predictive errors. The overview is limited to examples and by no means a full coverage of the available literature.

\subsubsection{Model and Parametric Uncertainty: Bayesian Approaches}

Bayesian uncertainty analysis has successfully been employed for parametric uncertainty quantification and propagation in MD simulations,\cite{Angelikopoulos2012} in rare events simulations,\cite{Chen2015,Hadjidoukas2015,Trendelkamp-Schroer2015,Prinz2011} in turbulence modelling,\cite{Cheung2011} in differential equation solvers, \cite{Koutsourelakis2009} or when modelleling alloys with surrogate models.\cite{Kristensen2014} Bayesian methods are widely used, despite having been questioned recently,\cite{Hand2014} for (knowledge based) parameter or model uncertainty estimations.\cite{Ghanem2006,Fossgaard2006,Nam2012,Metzner2010,VonToussaint2011,Bratholm2015}  In combination with experimental techniques they also helped to predict uncertainties in spectroscopic measurements\cite{Chodera2010} and in force field calibration.\cite{Cailliez2011}

Bayesian methods have an attractive internal mathematical coherence, formally linking the prediction with the observed data based on Bayes theorem.\cite{Karniadakis2006} 
Despite the prior information is subjective, the advantage of Bayesian analysis is its transparency with respect to biases and, thus, its wide applicability.
The ultimate aim is to say something about the real world with computer simulations based on models of reality. Thus, other methods to quantify uncertainty may be more appropriate for some questions. Especially, when keeping in mind that there is never a perfect model of the reality of the world.

\subsubsection{Knowledge Based Uncertainties: Bootstrapping and Filtering}

Methods dealing with knowledge based uncertainties which arise through finite sampling or partial observation of a system are bootstrapping and filtering. In bootstrapping an approximate sampling distribution provides us with insights about the uncertainty of a complex probability process.
Data are sampled from good estimates of a probability process model and then treated as if they were from the exact model;\cite{Nicholls2014} e.g., to account for the finite sampling error in time-series analysis\cite{Taylor2015} or in Markov dynamics.\cite{Metzner2009}
Through filtering the best statistical estimate of a natural system is obtained if only partial observation is possible. For example, extended Kalman filters can systematically correct for both multiplicative and additive biases in stochastic parameter estimation filtering.\cite{Gershgorin2010}

\subsubsection{Model Uncertainties: Implicit and Explicit Approaches}

Explicit stochastic models or surrogate models can make uncertainties, which are due to variability, explicit in the model specification.\cite{Sahlin2015}
To assess non-parametric model uncertainties in stochastic engineering systems, perturbation methods are suited for problems with small perturbations or fluctuations. Here, the stochastic quantities are expanded around their mean value via Taylor series or the inverse of a stochastic operator in a Neumann series.\cite{Karniadakis2006}
The polynomial chaos expansion (PCE) and its variants, provide a high-order hierarchical representation of stochastic processes, similar to spectral expansions. 
The PCE as well as Karhunen--Lo\`eve expansions are useful to quantify variability based uncertainties in coefficients and parameters.\cite{Haasdonk2013} Example applications of the PCE are the quantification of uncertainties in porous media flow, with better convergence than straightforward Monte-Carlo methods,\cite{Burger2014,Kroker2015}
and the construction of error bars in three-dimensional heat transfer problems.\cite{Karniadakis2006}
PCEs have also been used to  assess the predictive accuracy of stochastic models whose errors are due to limited data.\cite{Ghanem2006}
Many models in biomolecular solvation, porous media flow, and biochemical reaction networks need reduction to surrogate models due to their complexity to facilitate uncertainty quantifications of the full model.\cite{Waldherr2012,Lei2015,Chen2015}
An explicit approach to incorporate uncertainties into models described by (hyperbolic) PDEs is the addition of a random component, e.g., a time dependent coefficient modelled by the Ornstein-Uhlenbeck process or a random field coefficient with a given covariance in space.\cite{Haasdonk2013,Barth2016}

\subsubsection{Parametric Uncertainties: Sensitivity Analysis and Target Intervals}
In metal catalysis, sensitivity analysis simulations can refer to the concept of degree of structure sensitivity of a catalytic reaction with respect to activity, given as turn over rate or selectivity.\cite{Noerskov2008}
Here, we mean sensitivity of a numerical method with respect to small variations in the initial conditions and small errors parameters, which can result in chaotic behaviour, see Sec.~\ref{sec:chaosnum}.
Variability based uncertainty in model parameters has a local and a global component.

The separate investigation of the influence of each input parameter on the output is called local sensitivity analysis, which can be measured by computing the partial derivatives of an output function with respect to its input parameters. Global or correlated sensitivity analysis takes the entire design space into account to assess how the variability of the output parameters is effected by the variability in each input parameter.\cite{Fender2014} 
A way of dealing with parametric and measurement uncertainties is to propagate some uncertainties of the input variables through the studied system,\cite{Fusi2016}
e.g., with stochastic collocation methods as in peridynamics for crack propagation.\cite{Griebel2015}
Global sensitivity analysis of ordinary differential equations can also be performed with error-controlled PDEs describing the evolution of the probability density function associated with the input uncertainty.\cite{Weisse2011}
The Fisher information matrix is a useful sensitivity measure for stochastic dynamics in cases with a high number of parameters where straightforward, gradient based methods, are impractical.\cite{Dupuis2016,Tsourtis2015} 
For stochastic reaction networks (of complex biological phenomena) a reduced-variance, finite-difference, gradient-type sensitivity approach relying on stochastic coupling techniques for variance reduction is possible.\cite{Arampatzis2015}
Further measures in global sensitivity analysis are regression coefficients, Pearson correlation coefficients, Spearman correlation coefficients, Sobol indices, or Morris' elementary effects method for multidimensional functions as in finite element simulations.\cite{Fender2014,Idrisi2014}

In micro-kinetic models of chemical systems, the uncertainty in kinetic parameters refers to pre-exponentials, bond indices,\cite{Ulissi2011} or energies to determine reaction rates.\cite{Meskine2009} They are all prone to carry errors from previous calculations of parameters into the newly designed model which can be measured through the concept of 'degree of rate control'.\cite{Meskine2009} 
Another strategy to deal with parametric uncertainties uses target intervals instead of target values when searching for the optimal function to describe a problem, which ensures robustness against unintended variations.\cite{Fender2014} Confidence intervals, however, only describe a fraction of the range of uncertainty. A full characterisation of uncertainty would mean to provide the likelihood of all possible values in the interval.\cite{Sahlin2015} Nonetheless, many tables and two-dimensional graphs in publications in science and engineering express uncertainty or provide error estimates in form of (confidence) intervals. 
For parameters which are uncertain due to a lack of knowledge, often through idealisations or simplifications in the modelling process, fuzzy sensitivity measures might be a better choice than stochastic measures for the uncertainty analysis, because the assumption of infinitesimal deviations from the nominal system is not valid in such cases.\cite{Walz2015}

\subsubsection{Uncertainty Visualisation}

Visualisation of data is the window through which scientists examine their data for deriving scientific conclusions, and the lens used to view modelling and discretisation interactions within their simulations, see Fig.~\ref{fig:uncertainties}. 
Different visualisation approaches can lead to different interpretation of data sets and different scientific conclusions.
In some cases, the uncertainty in the visualisation can lead to wrong interpretations.\cite{Eklund2016}
Evidently, uncertainty is an important part of the information that has to be represented in order to avoid erroneous interpretation of the data. Nevertheless, the majority of two-dimensional and three-dimensional visualisation methods ignore errors and uncertainties of data, just as how they are propagating through the different stages of data analysis.
Uncertainty visualisation deals with uncertain data from simulations or sampled data, uncertainty due to the mathematical processes operating on the data, and uncertainty in the visual representations.\cite{Hansen2014}
Recently reported approaches include uncertainty visualisation in scalar fields,\cite{Pothkow2011,Pfaffelmoser2011,Pfaffelmoser2012,Poethkow2013,Pfaffelmoser2013,Pfaffelmoser2013a,Mihai2014} vector field ensembles,\cite{Ferstl2016}
animation methods to convey uncertainty in the rendering of volumetric data,\cite{Lundstrom2007} image segmentation algorithms,\cite{Al-Taie2014} and parameter space analysis.\cite{Heinzl2014a,Luboschik2014}
An \emph{ad hoc} uncertainty assessment of visualisation parameters, e.g., in radiology, is possible through considering several possible visualisation outcomes.\cite{Lundstrom2007}

Good visualisations provide reliable and unbiased communication to avoid missed discoveries, miscommunications, and, at worst, creating a bias towards the research that is easiest to display.
Especially, when dealing with complex subject matters good visualisation tools guiding the process of decision making and judgment of the computational results are important.\cite{McInerny2014}
When uncertainty visualisation is included in the decision making process final decisions can be taken with fewer reconsiderations.\cite{Riveiro2014} 
Therefore, especially in combination with enhanced three-dimensional environments for virtual reality,\cite{Norrby2015} the integration of uncertainty visualisation tools into computational chemistry, physics, and engineering would add tremendous benefits for multiscale modelling.
Apart from new perspectives to deeper understandings of molecules in drug discovery and materials design such tools can prevent the users from running into time consuming pitfalls and lower the knowledge barriers for making predictions of chemical systems. Moreover, they would allow to interactively gain deeper insights into simulation results by providing means for quantitative and qualitative analysis.

\subsection{Inspirations From Numerical Mathematics}
\label{sec:nummath}

The field of numerical mathematics and analysis is specialised in dealing with the errors and uncertainties in physical models, their implementation, and the predicted outcomes in actual simulations, Fig.~\ref{fig:uncertainties} (c) to (e).
Numerical error estimations are well established in fields such as bridge engineering, weather predictions, and finance where risk assessment is crucial to decision making.\cite{Deuflhard2003}

\subsubsection{Chaos and Numerical Stability Analysis}
\label{sec:chaosnum}

\begin{figure}[htb]
\centering
\includegraphics{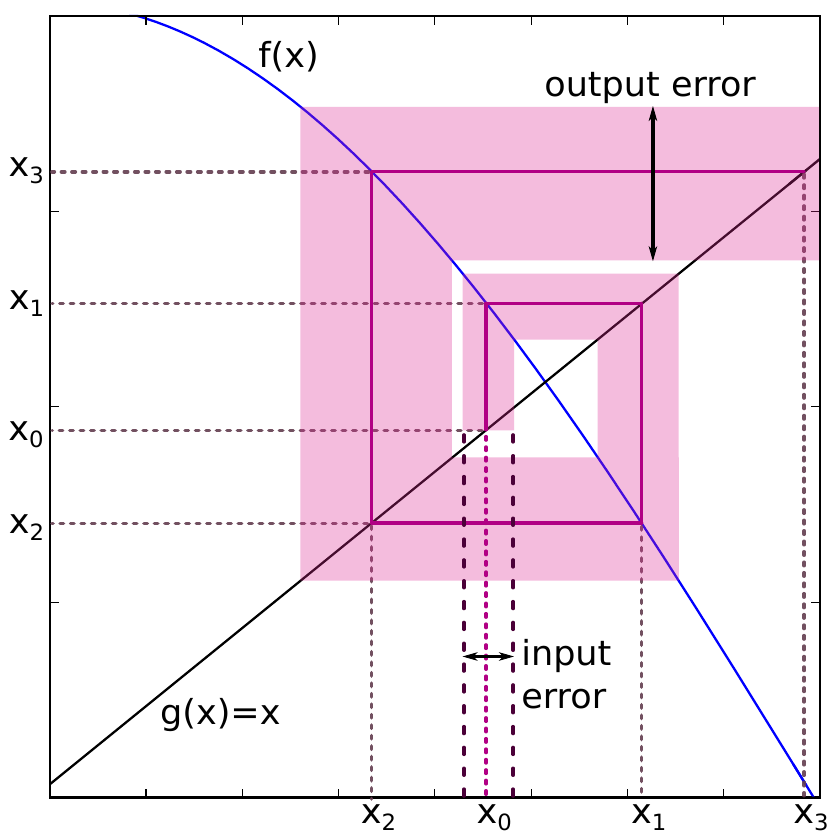}
\caption{Unstable iterative solver diverging from a repelling fixed point with input error enhancement.}
\label{fig:repel}
\end{figure}

In a classical mechanical systems chaotic behaviour is defined in terms of sensitive dependence on a few initial parameters. 
If the initial condition in a chaotic, non-linear model is uncertain, e.g., through small variations in parameters, then this uncertainty will evolve non-linearly, potentially rendering stable predictions impossible. Small uncertainties in the current state of the system will grow exponentially on average, see Fig.~\ref{fig:repel}. Uncertainty in the initial condition limits the utility of single forecasts. However, in many systems there are locally ergodic regions which are independent from the initial conditions and allow stable predictions. For example, local equilibria on the energy landscape of a chemical system or the vibrational motion of simple quantum systems.\cite{Jansen2012,Back2004}

Dynamical systems governed by Hamilton's equations can exhibit chaos;\cite{Almeida1990}
unsurprisingly, the linear TDSE can as well.\cite{Kubotani2006,Chakraborty2015} 
Chaos becomes even more important in cases where the non-linear SE is required to capture the essential physics, as in band gap design, ultra cold molecules,\cite{Carr2009} or in magnetic thin films.\cite{Anderson2014}
Even in Transition State Theory, \cite{Wigner1938,Wigner1939,Laidler1983} a common approach to determine reaction rates, some chemical many-body systems can show chaotic behaviour.\cite{Prigogine2002}

Numerical computations are performed with numbers of limited accuracy due to all computers using finite and discrete machine representations (hardware uncertainty), Fig.~\ref{fig:uncertainties}. Rounding errors, due to precision arithmetic, are present. They become problematic when significant digits are eliminated during a subtraction, when inaccuracies are amplified while summing up a large amount of very small numbers or while dividing by a small number.\cite{Deuflhard2003,essex2000} Then the numerical algorithm may become unstable and sensitive to small changes in its intermediate results, while at the same time -- similar to chaotic behaviour -- loosing the ability to be predictive. 

Sources of instability may also originate from the discretised equations of the numerical solver. 
Many systems described by ordinary differential equations, like Newton's equations of motion in MD,\cite{Alder1959,Rahman1964} or by PDEs, like TISE and TIDE,\cite{Schroedinger1926} require the often tedious choice of a temporal or spatial discretisation step size for an integrator.
For efficient MD simulations the step size should be chosen as large as possible, whereas too large step sizes lead to instabilities, i.e., blow up or drift off the sought solutions.\cite{Davidchack2010}

A standard approach to tackle the problem is the forward-backward sampling for stability analysis of MD solutions. 
Still, such stability analysis only gives insights about the behaviour of the error propagation, the quantitative size of the numerical errors  stays unknown and uncontrolled.
If modellers are unaware of the underlying numerical errors and their influence on the computed results, they may be prone to give physical meaning to artefacts originating from too large discretisation steps. Pressure profiles in spatially inhomogeneous systems not being uniform could be misinterpreted as evidence for the presence of some hidden internal forces within the system.\cite{Davidchack2010}
Similarly, in PDE-constrained optimisation numerical errors in discrete solutions can pollute a quantitative measure of the objective function such that the optimised design reflects the errors in the discretisation rather than the physics of the problem.\cite{Hicken2014}

Thus, to ensure good quality of the simulation algorithms both, numerical stability analysis and a mathematical proof of convergence to the sought solutions, are recommended and common in many research areas. The best cases to prove convergence are explicit and verifiable error estimators of the form $\| X_{\text{numerical}}-X_{\text{exact}}\| \leq C$, where the exact solution $X_{\text{exact}}$ is unknown and the constant $C$ is the desired error tolerance of the numerical method.

The goal of a numerical stability analysis is: (i) to assess how the tiny intermediate rounding errors of floating point arithmetics affect the final calculation result and (ii) to quantify how uncertainties are introduced or amplified by the various subtasks of a calculation.
In a stable numerical method the output error is less or equal to the input error and small variations in the initial conditions only have a small impact on the final result. Each change in an algorithm requires a careful analysis to keep chaos under control.

\begin{figure}[htb]
\centering
\includegraphics{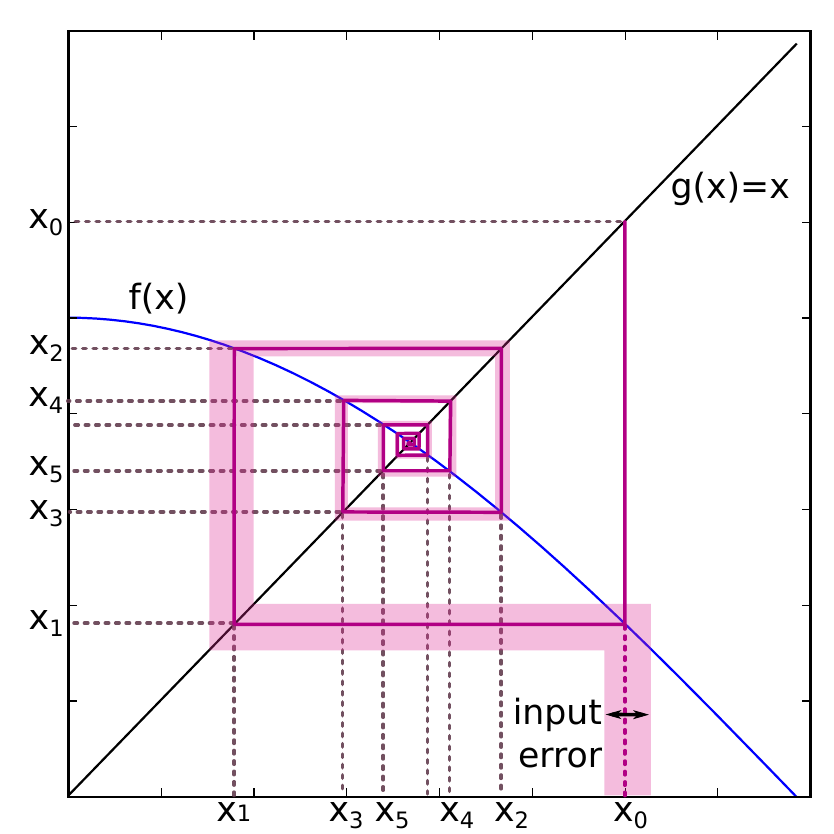}
\caption{Stable iterative solver converging to an attractive fixed point.}
\label{fig:attract}
\end{figure}

As example we consider fixed point iterations, see Fig.~\ref{fig:attract}, of a continuous function to find $f(x)=x$ through an iterative discrete numerical method $x_{n+1}=f(x_n), \, n=0, 1, 2, \dots$. 
The self-consistent field or HF method\cite{Hartree1929,Fock1930} for electronic structure calculations, is a mean field approach and the original iterative numerical solver a fixed point iteration to find a balanced charge distribution in the field. Similarly, the 
Newton-Raphson method to find roots of a function $f(x)=0$ with $x\in\mathbb{R}$ of a continuous function $f:\mathbb{R}\rightarrow \mathbb{R}$ by iterating
\begin{equation}
x_{n+1}= h(x_n)=x_n-\frac{f(x_n)}{f'(x_n)}, \quad n=0, 1, 2, \dots
\end{equation}
is interpretable as a fixed point iteration to find $h(x_n)=x_n$.\cite{Susanto2009}

The fixed point in Fig.~\ref{fig:attract} is an attractive fixed point and the iteration converges while the input error contracts. In contrast, a small variation in the pre-factor of the function $f(x)$ leads to the repelling fixed point in Fig.~\ref{fig:repel}. Then the initial error is amplified in an unstable, diverging numerical procedure, where small changes in the initial input hugely impact final results.

\begin{figure}[h]
\caption{The seven roots of $f(z)=z^7-1$, $z\in\mathbb{C}$ as a function of the initial  value $z_0$. The seven basins of attractions are the areas of stability containing the fixed points to which the Newton-Raphson root finding solver converges to. The areas between the basins are potentially unstable areas.}
\centering
\includegraphics{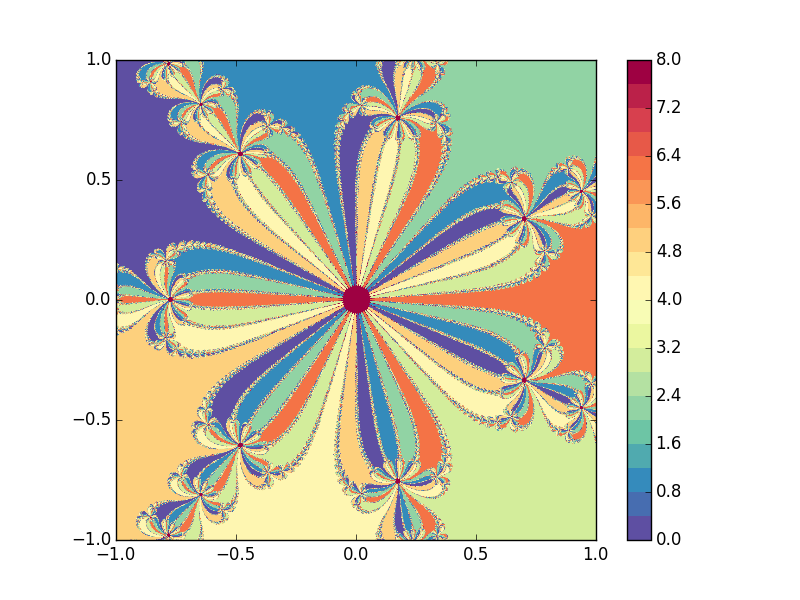}
\label{fig:stabarea}
\end{figure}

Each algorithm has a specific window of stability and predictiveness.
Fig.~\ref{fig:stabarea} shows the basins of attractions or areas of stability for the convergence of the Newton method while finding the roots of $f(z)=z^7-1$, $z\in\mathbb{C}$. The larger basins of attraction are separated by small areas where only small changes in the initial value cause the iteration to converge to a different final solution.

\begin{figure}[h]
\caption{Number of iterations until the Newton-Raphson method converges to one of the roots of $f(z)=z^7-1$, $z\in\mathbb{C}$ as a function of the initial value $z_0$.}
\centering
\includegraphics{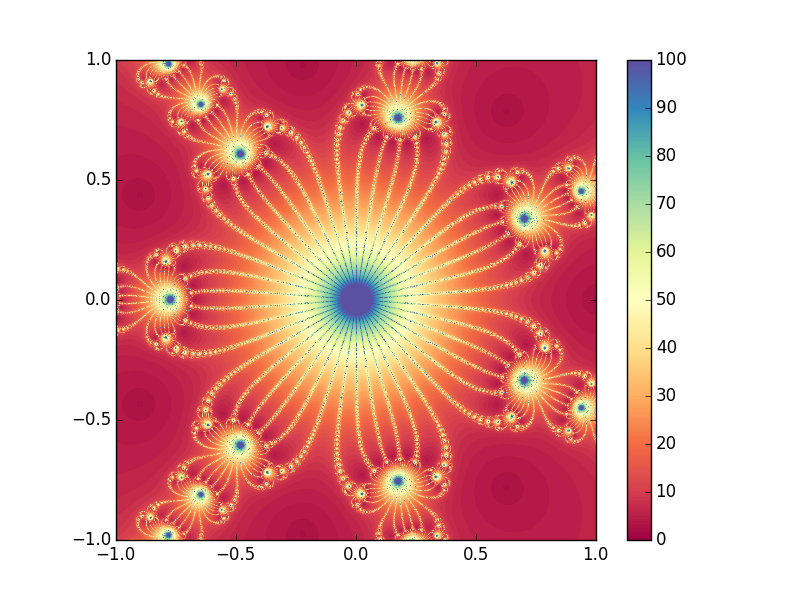}
\label{fig:convspeed}
\end{figure}

Therefore, judging an iterative algorithm only based on a couple of example calculations and a comparison of the resulting numbers, as often done in chemical accuracy analysis, is insufficient. The whole range an algorithm can cover requires consideration and analysis with respect to stability and limitations of predictiveness.
The number of steps an algorithm takes to converge to a solution may vary depending on the initial positions, see Fig.~\ref{fig:convspeed}. One algorithm may be slower compared to another for the same initial value, but exhibit the same speed of convergence when a different starting point is used. Here, mathematical convergence proofs using the language of sequences, in contrast to looking at special cases, provide the advantage of gaining general insights and clarity about where limitations might occur.
Additionally, in each iteration step one algorithm might perform only one floating point operation, whereas the other calculates several complex and computationally more demanding intermediate steps. Obviously, the computing time to convergence is much slower for the second algorithm. The concept of time complexity of an algorithm, connecting the input dimensionality of a problem with the number of elementary operations performed, would be better suited to judge algorithms than the number of iteration steps. 

In the field of numerical analysis, an area of applied mathematics, special interest lies on the construction of (a) efficient and fast solvers which are (b) robust in case of disturbances for example due to measurement and numerical precision errors or modelling uncertainties. Additionally, a solid and reliable solver includes a (c) mathematically provable and verifiable error estimate. In particular, (b) and (c) ensure that chaotic behaviour is avoided and the numerical method is predictive. The field of numerical mathematics provides many concepts to deal with numerical errors, model reduction, and structural uncertainties.
The quality of numerical methods is judged with respect to their convergence towards the exact solution, their velocity of convergence, the effort and computational complexity they require, and the robustness of their solution.
To avoid the misinterpretation of numerical artefacts as physical effects a clear separation of the analysis of the errors of numerical methods and the physical models is recommended.
Further, there are several sources of numerical errors: discretisation errors, statistical errors of short time trajectories, truncation and precision errors, etc.
By investigating their impacts separately,  e.g., on averages computed from MD trajectories, via intrinsic or extrinsic error estimates and stability analyses of the numerical methods, they can be quantified and systematically improved.

\subsubsection{Error Estimation Techniques and Convergence Analysis}

Quality measures can be assigned extrinsically by comparing errors of the final results without code modifications, or intrinsically, by designing methods which already allow to estimate the error in the predicted outcome a-priori or a-posteriori.
Extrinsic methods are common in computational chemistry. Intrinsic methods are challenging to design and inspirations from other fields could fundamentally change the chemical modelling process.

The presentation of elaborate error estimates, whether reasoned or derived, is rare in the field of computational chemistry; perhaps it is due to the mistaken belief in the exactness of theory.\cite{Nicholls2014a} Still, we will give an overview of a couple of examples.

Based on extrinsic analysis single floating point precision was claimed to be sufficient for calculations after designing algorithms and methods with double precision.\cite{Vysotskiy2011} Extrinsic stability analysis of quantum chemistry algorithms based on the introduction of random noise at the order of magnitude of floating point precision, however, showed that even well established techniques in quantum chemistry can have serious defects and proved the neglect of double precision arithmetics wrong.\cite{Knizia2011} 

Further, the effects of numerical discretisation errors on computed averages in MD simulations,\cite{Davidchack2010}
upper bounds for the approximation errors of long-time statistical dynamics with Markov models of molecular kinetics,\cite{Prinz2011a} and quantitative probabilistic error estimates for excited state calculations in Born-Oppenheimer MD have been investigated.\cite{Bayer2013}

A priori error estimates are possible for both linear TIDE and TISE as well as non-linear KS-DFT equations whose solvers use simplified atomic orbitals.\cite{Chen2015a,Chen2015b} Simplified atomic orbitals, often with polynomial-type and confined Hydrogen-like radial basis functions, are highly efficient in electronic structure calculations since they require less basis functions to yield similar precision compared to other discretisations. The related methods, in contrast to plane-wave or real-space-grid basis set methods, lack, however, systematic convergence and a posteriori error estimates since the increase of basis sets can be achieved in several ways.\cite{Chen2015a} 
Explicit error estimates, existence, and uniqueness of solutions comparing to the full TISE have been mathematically proven for the CC method.\cite{Rohwedder2013,Rohwedder2013a}
In some cases, the error control of electronic structure solution is present, however, only on one level of the set of underlying equations, e.g., in the approximation the electron repulsion integrals.\cite{Aquilante2008,Aquilante2009}
Finally, a model reduction error in tensor network approaches for open quantum many-body systems can be estimated.\cite{Werner2016}

Since TISE and TDSE are PDEs, we also mention some references for PDE solvers.
Reliable and efficient a priori or a posteriori error estimators are available for numerical solvers of linear parabolic or parametrised PDEs,\cite{Haasdonk2011,Urban2013,Dumbser2016}
PDEs with stochastic influence,\cite{Haasdonk2013} linear elliptic PDEs,\cite{Sen2006} 
and for optimal control problems with a PDE constraints.\cite{Leugering2012,Hicken2014}
Convergence studies and stability analysis of numerical schemes are available for random PDE solvers.\cite{Barth2016}
A space-time adaptive wavelet Galerkin method to solve parabolic PDEs, similar to the CC method mentioned above, was tested for the convergence rate to the exact solution and applied to a diffusion-convection reaction.\cite{Kestler2015}
In reaction-diffusion master equations, local error estimates can be constructed through operator splitting. They allow for an adaptive control of the time step discretisation to take as large as possible steps maintaining a certain accuracy.\cite{Hellander2014}

As mentioned above, some current approaches in computational chemistry include extrinsic methods to analyse the stability of quantum chemical simulation codes and algorithms. Still, there is only little information to be obtained from these approaches about uncertainties and error bars. Some approaches focus only on the error of one subtask of the calculation leaving the question of the clear global numerical error control open. A better strategy would be to design methods and models whose results intrinsically include error bars, uncertainty information, stability estimates and, thus, a clear interrelation of numerical precision and chemical accuracy. Such methods steer away from intuitive judgements of results and provide precise control over how much numerical precision is required for a certain accuracy in the predicted outcomes.
To reach for such methods, new concepts and mathematical descriptions specially targeted at nanoscale engineering based on measurements and simulations beyond the SE, Heisenberg's matrix mechanics, or Feynman's path integral formulation are required. 
Making Feynman's path integral formalism mathematically meaningful is already quite tough,\cite{Johnson2000} proving error bars after discretisation of the involved infinite integrals close to impossible.
We believe that better tools, methods, and strategies for numerical error analysis are required in computational chemistry. To reduce the risk of wrong interpretations we recommend both extrinsic, e.g., statistical analysis after the implementation of the code, and intrinsic approaches with models, code, and numerical methods containing explicit error and uncertainty estimates for a thorough validation and verification procedure.

\subsection{Quality Control: Validation and Verification}
\label{sec:valiveri}

\begin{figure}[htb]
\centering
\def\svgwidth{8cm}
\sffamily
\footnotesize 
\includegraphics{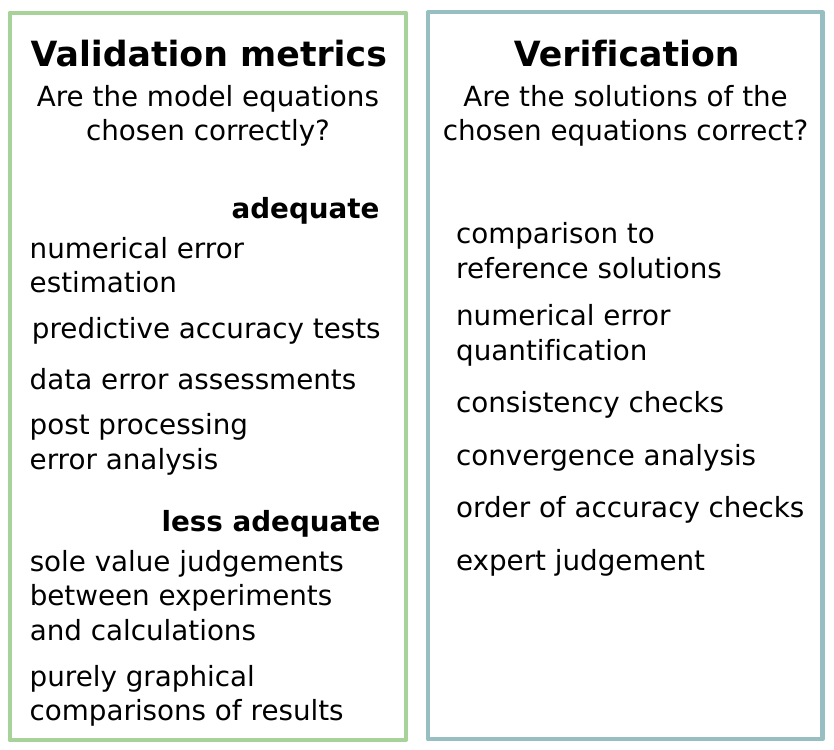}
\caption{Validation and verification techniques.}
\label{fig:modcycle}
\end{figure}

Verification and validation are an integral part of the modelling cycle to ensure reliable and reproducible simulation results, see Figures~\ref{fig:modcycletot} and~\ref{fig:modcycle}.
Verification is the process by which we ensure that the algorithms have been implemented correctly and that the numerical solution approaches the exact solution and that  the equations of a model are correctly solved. 
Validation on the other hand is primarily concerned with the correct choice of model equations. 
In some cases, assumptions about the potential energy landscape of chemicals from different electronic structure models have been misleading due to lack of validation.\cite{Viegas2014}
The aim of model validation is to provide reproducible outcomes so that researchers can reach the same conclusions, when given the same data and analysis tools. 
The fundamental strategy of model and solution validation is to assess how accurate computational results are compared to  simulated physical phenomenon with quantified error and uncertainty estimates for both.\cite{Oberkampf2002} Thus, appropriate measures to validate the assessment of uncertainties are important.
Equally, visualisation should explicitly be considered as part of validation and verification in the process of predicting and deciding based on models.\cite{Kirby2008} 

Code and solution verification procedures in computational simulations include several tests: expert judgment, error quantification, consistency checks, convergence analysis of the discretised problem to the exact solution, and order of accuracy tests, which determine whether a discretisation error is reduced at the expected rate during consistent refinement of discretisation meshes.\cite{Oberkampf2002,Roy2005,Rider2016}
Among these tests the least rigorous is expert judgment, where an expert is given the code output for a given problem and asked to determine whether or not the results appear to be correct. 
Error quantification tests involve quantitative assessment by comparing numerical solutions with exact or benchmark solutions. Subsequently, the error is judged for whether it is sufficiently small. The most rigorous code verification is the order of accuracy test.\cite{Roy2005} Mesh refinement studies are common to estimate discretisation errors, despite their imperfections.\cite{Hicken2014,Mosby2016}
Code independent, external verification analysis often uses reference solutions,\cite{Ingraham2013} 
which are usually taken from well studied, known, and understood systems. 

The 'What is your favourite DFT functional' popularity contest is a form of expert judgement which gives immediate warnings for unreliable developments and includes an accountability for human mistakes. However, such polls lack quantitative measures of errors and uncertainties in the simulation methods and, thus, they are less concerned with progressing accuracy or rigour in the simulation method itself.\cite{Swart2016}

Qualitative validation by comparing results of simulations and experiments graphically, despite being quite common, gives little quantitative indication of how the agreements between computational results and experimental data vary over the range of an independent variable -- spatial coordinates or time -- and how uncertainties and errors contribute.
Essential inputs for good validation are well characterised experimental results and a metric capturing the subtleties of the differences that are detected in the comparisons to give clear hints about where improvements are recommended. 
Many properties of chemical systems can already be predicted realistically and provide good starting points to validate new developments: spectra, energy differences (difficult for clusters), excitation energies, and atomistic structures.

Validation metrics are computable measures that allow to quantitatively compare computational and experimental results over a range of input or control variables for a sharper assessment of computational accuracy.\cite{Oberkampf2006} 
Good validation metrics include (a) estimates of numerical errors, (b) quantitative evaluations of the predictive accuracy of the simulated quantity of interest inclusive all modelling assumptions and physical approximations, (c) estimates of the errors resulting from the post-processing of the experimental data, and (d) they explicitly incorporate an estimate of the measurement errors in the experimental data.\cite{Oberkampf2006}

Inadequate metrics, despite often applied  in the field of computational chemistry, are value judgements between experimental and computational results. Here, computational results are judged as being adequate because they lie within the uncertainty band of the experimental measurements.\cite{Oberkampf2006}
Chemical accuracy often represents such value judgements and has limited meaning with respect to quantitative accuracy and predictiveness. Examples include: the validation of CC F12 with respect to atomisation energies\cite{Klopper2010} and the validation of enthalpies from the correlation consistent composite approach (ccCA)\cite{DeYonker2006,Weber2015}. Many other TDSE and TISE solvers have been assessed by comparing mean deviation, mean absolute deviation, standard deviation, or maximum absolute deviation of the calculated results, e.g., bond distances, with experiments. \cite{Coriani2005}
In cases where experimental results for comparisons are lacking, many methods in computational chemistry have been judged based on value comparisons with other simulations, whose results are assumed to lie within chemical accuracy.\cite{Boese2013,Boese2015,Korona2013,Dral2015,Ramakrishnan2015} 

Reduced models in quantum chemistry are often validated through comparison with models whose model is thought to be closer to the physical reality. For example, the method of increments truncation to calculate correlation energies\cite{PAULUS2007} was compared with \emph{ab initio} DMRG benchmarks.\cite{Fertitta2015}

An alternative strategy of quality measurement is to consider measurements specifically tailored to the requirements of simulations. 
Aerospace simulations and modelling generally use well-established inputs that are measured with relatively high accuracy and relatively small variance,\cite{Woltosz2012} 
whereas chemical experiments are often performed without any simulation in mind.\cite{Boyce2015} 
Recent advances in measuring theoretical quantities directly instead of using numerical approximations or intermediate models include: 
the prediction of protein assemblies from cryo-electron microscopy data,\cite{DeVries2016} the extraction of time-averaged equilibrium probability distributions for an individual molecule via tunnelling microscopy,\cite{Palma2015} or the direct construction of energy landscapes\cite{Laidler1983} from experimental single-molecule time-series.\cite{Li2008,Sultana2013,Taylor2015}
The mechanical forces between two covalently bound atoms can now be measured and, thus, the forces can act as control parameters to design potential energy landscapes,\cite{Wales2006} rather than being structural constraints.\cite{Ribas-Arino2009}
By putting more effort, time, and expenses into the measurement stage of the modelling cycle, physical and chemical sciences can gain detailed knowledge about risks and uncertainties in large and complex systems.

\section{Challenges for the Future of Computational Chemistry}
\label{sec:future}

The huge progresses in quantum chemistry over the past decades have been driven by the development of computers with increased hardware power and the implementation or adaptation of efficient algorithms.\cite{Deglmann2015,Thiel2011} In future, solutions to grow and advance the field, especially in the context of multiscale and multi-physics simulations, will not only rely on greater computing power, but primarily in theoretical, conceptual, and algorithmic work across disciplines.\cite{Ramasubramaniam2011}
In industrial environments with time and cost constraints, e.g., in terms of man and compute powers, any computational approach to a specific question will always represent a compromise between accuracy and expected computational effort.\cite{Deglmann2015}
Traditionally, computations -- also with quantum chemical models -- are performed to gain better understanding in cases where experimental data are difficult to interpret, thereby bridging between theoretical and experimental observations.
Recently, the future of computational and theoretical chemistry was envisioned with an increased importance for experiments: by reliably predicting the design of experiments or candidates for chemical synthesis and by providing independent information and insights through calculations,\cite{Thiel2011}
even before any experiments have been conducted.
Experimentalists will be able to unlock their creativity and interactively play with their ideas of what to study in simulation tools. They will be able to run fast pre-screening tests as well as prioritise and improve experimental design while at the same time having confidence about the reliability of their predictive model including safeguards as to its validity or meaningfulness. Moreover, scientific computing could become a valid tool for targeted research development and for safety assessment of human pharmaceutical compounds, drug induced toxicities,\cite{Huang2016} or airplane designs.

The philosophical question whether computer simulations will replace experiments can be answered the following way: only experiments can generate new empirical data, can operate directly on the target systems, and only they can be employed to test fundamental hypothesis. Consequently, computer simulations cannot completely take over their distinct epistemic role in science.\cite{Duran2013}
Compared to experimental techniques, simulations have similar advantages and caveats which makes them equally valid as long as they are used where appropriate and their limitations are clearly stated.\cite{Gray2015} 
Through the establishment of reliable protocols and awareness of caveats, the field of computational (bio-)chemistry has started to become an equal partner to experimental research. 
\cite{Schlick2011,Gray2015}
However, to be recognised as a standard technology in chemistry similar to NMR probings their predictive power needs to be proven. To predict outcomes before doing any testing in the lab requires wide support for the judgement of the quality of the results and the final decision making, as discussed in the previous sections.

To reach reliable, robust, and accurate predictions for \emph{in silico} engineering of new materials and drugs, with black box programs,\cite{Lopez2012} several challenges need to be overcome.
One of the hurdles is the reliable automation of simulation methods for virtual testing and hypothesis probing, where experiments are not feasible.
Here, the challenge is to give the user confidence about potential failure and the entire process of virtual testing.
To enable researchers in chemistry labs to use computational modelling before they do an experiment, we need a huge variety of reliable data- and equation-based models to describe their problems.
Further, management and control of interacting uncertainties and risks connected to the whole design and simulation process, from the quality of input data, over model construction, and, finally, in the process of judging the results, is essential.
The communication of numerical, human, and modelling errors can ensure a rigorous judgement and decision making process based on the predicted results. Systematic error estimates will lead to quantitative comparisons of simulation results. 
Transparency throughout the entire modelling cycle is central to improve reproducibility. 
When combinations of models or methods are used in the simulations, the error assessments require clear pictures and separation of various sources of uncertainty: from parametric uncertainties over structural model uncertainties to numerical uncertainties and visualisation uncertainties in final data analysis.

In addition, the underlying assumptions, strengths, and limitations of available models and parameters need clear communication to all users of a simulation approach, although the user might have limited knowledge about the underlying process of model design. Here, new statistical tests and visual tools with novel concepts to communicate assumptions, risks, and uncertainties would be of great support.
Such new concepts can increase the reproducibility and reusability of case specific models which are quite common in the context of biochemistry.

The initial expenses and investments when separating the uncertainty and error assessments in the processes of model creation, reduction, and numerical implementation, pays off when validation beyond expert judgement becomes possible.
Tools to manage uncertainties in the modelling cycle are more than an assemblage of existing tools, since they usually lack natural interfaces to manage and quantify uncertainties in physical phenomena.\cite{Panchal2013}
Based on the error and uncertainty estimation intrinsic and extrinsic to simulation methods and models, uncertainties and risk can be made visible and used for guidance in decision making processes. Another form of risk reduction is to design measurement setups specifically with simulations in mind and to find concepts relating measurable properties with complex properties.
Finally, the accurate \emph{in silico} chemical risk assessment is a statistical issue. \cite{Huang2016}
However, many mathematical formulations of models in computational chemistry, e.g., Feynman's path integral formalism, are very challenging when it comes to the explicit inclusion of uncertainty and error estimates.

In summary, we need systems that allow to communicate the context, the history, and the uncertainty of results in a generally comprehensive way.
Therefore, the design of tools which include error and uncertainty quantification across lengths and time scales requires new chemical, physical, mathematical, and visual concepts.
In future, computational modelling will first transition the chemistry related disciplines into targeted research environments through \emph{in silico} molecular systems engineering and then current theories will be challenged to test and expand qualitative chemical concepts on the basis of increasingly accurate computations.
The field of computational chemistry has the potential to become an equal partner to experiments with equal weight in decision making by moving from highly specialist method design to generally applicable tools accessible to researchers from various communities.
By addressing the open challenges discussed in this review we will move one step closer to the \emph{in silico} failure test of airplanes,  potentially replacing the construction of a whole test plane, since failure risks and uncertainties -- for example due to fatigue of materials under stress -- will be easier to integrate into larger physical scales when they are reliably assessed and predicted in molecular modelling.

\section*{Acknowledgements}
J.B.R. thanks the Alexander von Humboldt foundation for a Fedor Lynen Research Fellowship and the British Engineering and Physical Sciences Research Council (EPSRC) for a Knowledge Transfer Fellowship.

\bibliography{review,mypapers}

\providecommand{\latin}[1]{#1}
\providecommand*\mcitethebibliography{\thebibliography}
\csname @ifundefined\endcsname{endmcitethebibliography}
  {\let\endmcitethebibliography\endthebibliography}{}

\end{document}